\documentclass[12pt]{elsarticle}
\usepackage{graphicx}
\usepackage{amsfonts}
\usepackage{amssymb}
\usepackage{amstext}
\usepackage{amsmath}
\usepackage{listings}
\usepackage{float}
\usepackage{multicol}
\usepackage{natbib}

\journal{Nuclear and Instruments Methods in Physics Research Section A}
 
\begin{document}
\begin{frontmatter}

\title{FIPSER: Performance Study of a Readout Concept With Few Digitization Levels for Fast Signals}

\author[GTphys]{B. Limyansky\corref{cor1}}
\cortext[cor1]{brent.limyansky@gatech.edu}
\author[GTphys]{R. Reese\corref{cor2}}
\cortext[cor2]{bobbeyreese@gmail.com}
\author[GTee]{J.~D. Cressler}
\author[GTphys]{A.~N. Otte}
\author[GTphys]{I. Taboada}
\author[MSU]{C. Ulusoy}
\address[GTphys]{School of Physics and Center for Relativistic
  Astrophysics, Georgia Institute of Technology. Atlanta, USA}
\address[GTee]{School of Electrical and Computer Engineering, Georgia Institute of
  Technology. Atlanta, USA} 
\address[MSU]{Dept. of Electrical and Computer Engineering, Michigan
  State University. East Lansing, USA}

\begin{abstract}

We discuss the performance of a readout system, Fixed Pulse Shape Efficient Readout (FIPSER), to digitize signals from detectors with a fixed pulse shape. In this study we are mainly interested in the readout of fast photon detectors like photomultipliers or Silicon photomultipliers. But the concept can be equally applied to the digitization of other detector signals.
FIPSER is based on the 
flash analog to digital converter (FADC) concept, but has the potential to lower 
costs and power consumption by using an order of magnitude fewer discrete voltage levels.
Performance is bolstered by combining the discretized signal with the knowledge of the underlying 
pulse shape. Simulated FIPSER data was reconstructed with two
independent methods. One using a maximum 
likelihood method and the other using a modified $\chi^{2}$ test. 
Both methods show that utilizing 12 discrete voltage levels with a
sampling rate of 4 samples per full width half maximum (FWHM) of the pulse achieves an amplitude 
resolution that is better than the Poisson limit for photon-counting experiments. 
The time resolution achieved in this configuration ranges between $0.02 - 0.16$ FWHM and
depends on the pulse amplitude. In a situation where the waveform is composed of two 
consecutive pulses the pulses can be separated if they are at least $0.05 - 0.30 $ FWHM apart with an amplitude
resolution that is better than 20\%.  
\end{abstract}

\begin{keyword}
Particle astrophysics \sep Front end electronics \sep Photon counting
\sep Silicon photo-multipliers \sep Photo-multipliers

\end{keyword}

\end{frontmatter}

\section{Introduction}
\label{sec:intro}

	Thanks to new generations of instruments, 
	the past decade of astroparticle physics has been one of dramatic experimental 
	advancement. The success was in large part due to the scaling of previously proven 
	experimental techniques to achieve orders of magnitude higher sensitivities. 
	One of the consequences of the scaling is that instruments now use a much larger number of signal channels than before. 
	
	A typical astroparticle experiment employs several thousand channels, and the next 
	generation will utilize tens, if not hundreds, of thousands of
        channels. The shear increase in the number of channels is a
        potential challenge when 
	it comes to meeting power and cooling requirements at remote locations, not 
	to mention costs. Mitigating these challenges requires new concepts to digitize 
	signals that not only significantly lower the costs and power 
	per channel, but simultaneously increase the sampling rate and resolution. 
	
	The readout concept proposed here has applications in many fields beyond astroparticle physics and is based on FADCs. 
	Flash Analog to Digital Converters (FADCs) are a preferred means for signal digitization of any kind. 
	However, they are costly and consume a non-negligible amount of power ($>100 \; \text{mW}$), even when 
	running at modest sampling speeds of a few hundred megasamples per second and 8-bit resolution. For example the Analog Device AD 9283-100 samples with 100 MS/s at 8\,bit resolution and consumes 90\,mW, whereas the AD 9284 samples with 250 MS/s and consumes more than 300\,mW \cite{AD}.
	
	A good alternative in terms of cost and power that has become prevalent over the past two decades are switched 
	capacitor arrays (SCAs). Two prominent examples of SCAs are the TARGET chip \cite{gary1,gary2,gary3} and the Domino Ring Sampler (DRS) \cite{stefan1}, which are widely used in astroparticle and high-energy physics experiments \cite{gary1,stefan2,stefan3,stefan4}. The DRS4 chip, for example, consumes 17.5\,mW per channel when sampled at 2GS/s\footnote{This does not include the actual digitization which is only activated when a readout command is issued}.   In an SCA based system, unlike in an FADC system, 
	the signal amplitude is not digitized right away. Instead, the analog signal 
	values are stored in an array of capacitors and the stored analog values digitized when a readout 
	command is given. Common disadvantages of SCAs are limited trace lengths that can be digitized, complicated calibration, and deadtime.
		
	We propose a different approach to digitize signals in future astroparticle experiments, 
	called \textbf{F}ixed \textbf{P}ulse \textbf{S}hape \textbf{E}fficient \textbf{R}eadout (FIPSER). 
	FIPSER is based on the FADC concept, but allows power and costs savings that are similar 
	if not better than in the SCA approach. The major advantage of FIPSER over SCA based systems is the 
	continuous, dead time free digitization of analog signals without having to stop sampling, 
        which opens up new avenues to apply event selection at the detector's trigger level. As a 
	result, experiments utilizing FIPSER could potentially see a significant reduction in data volume while increasing the sensitivity of the instrument.

%

\section{FIPSER Concept}
\label{sec:concept} 

	In an FADC system, the input range is discretized into $2^{n}$ levels, where n-bit is the 
	resolution of the FADC. For example, an 8-bit system has $2^{8} = 256$ equally spaced levels.

	A typical FADC system is realized with $2^{n-1}$ comparators with the threshold of each one being 
	fixed to one of the $2^{n-1}$ discrete levels of the FADC. The signal that needs to be 
	digitized is fed to the input of the comparators, and the digitized output of the FADC is given
	by the comparator that triggers with the highest set threshold level. 
	
		\begin{figure*}[!ht]
			\centering
			\includegraphics [width=\textwidth]{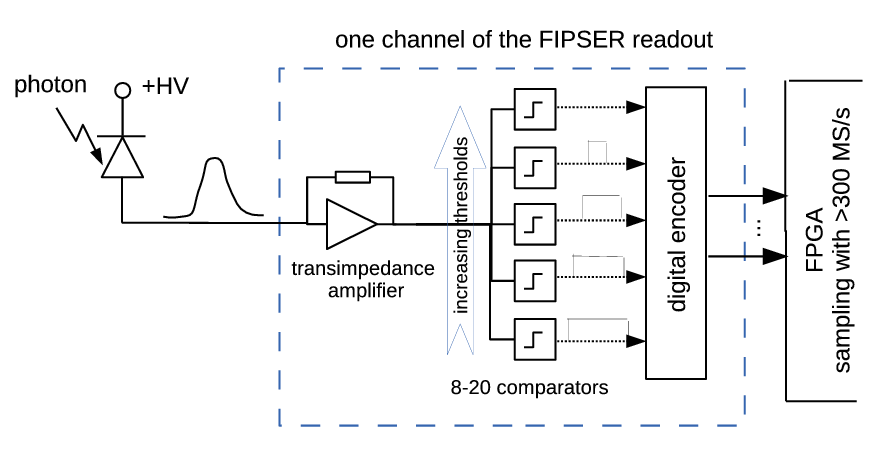}
			\caption{A possible implementation of FIPSER.}
			\label{fig:block}
		\end{figure*} 
	
	The power consumption and data volume produced by an FADC system is mostly determined 
	by the number of comparators. As such, reducing the amount 
	of comparators results in a significant power saving and smaller data volume. 
	
	The number of discriminators can be reduced in applications that do not require equal spacing
        of the discrete digital levels. The majority of astroparticle applications fall in this category because
        for most of them a fine spacing between the levels is necessary for the digitization of  
	small signal amplitudes but a coarse spacing is sufficient for signals with large amplitudes. 
        This is, for example, true if the recorded signal is subject to Poisson fluctuations. We designed FIPSER with the requirement to reconstruct a signal below the Poisson fluctuation limit and assuming that the shape of the signal is known.
	
	Figure~\ref{fig:block} shows a conceptual implementation of
        the FIPSER for the detection of photon signals from, for example, air showers or Cherenkov emission in water. 
        After an amplification stage 
        the analog signal is split into $n$-different branches that each lead to the input of a separate  
	comparator that is set to a unique threshold value. The output of a comparator 
	is logic high when the input signal is above the comparator's threshold value, and logic low otherwise. 
	The comparator outputs are connected to an FPGA via an optional digital encoder to reduce 
	the number of logic signal lines. The FPGA records the status of the input signals with 
	a fixed sampling rate, and further processes the data before it is recorded on hard disk. 
	Possibilities for online processing of the data with the FPGA include signal and time 
        extraction that could be used in advanced trigger algorithm that are executed by the FPGA
        as well.

        Other applications and implementations than the above discussed can certainly be thought of. 
        The objective of
        this paper is to demonstrate the feasibility of the FIPSER concept and 
        not to provide the technical details of a specific implementation.
	
	FIPSER can be described as being an FADC with varying resolution throughout the dynamic range. 
	As we will show in Section~\ref{sec:results}, as few as 12 comparators are sufficient to cover a dynamic 
	range of three orders of magnitude with an amplitude resolution that is better than the Poisson limit. 

        In the case when only one or two discriminator levels are used the proposed system is equivalent to many TDC readout systems that are used in a wide variety of experiments, e.g. \cite{cream,milagro}, which is not what we aim at with FIPSER. The studied system is better compared to the digitization of signals that are amplified with logarithmic amplifiers or detector signals with intrinsically logarithmic response \cite{porro}.
	
\section{Simulation of FIPSER}
\label{sec:config}

	To quantitatively assess the performance of FIPSER, we designed a case study reminiscent of typical situations
	encountered in astroparticle physics. For example, the chosen configuration could be used 
	to model the digitization of a signal produced by Cherenkov-light flashes, such as those 
	detected by the gamma-ray detectors HAWC~\cite{HAWC} and VERITAS~\cite{VERITAS},
	or by the IceCube neutrino detector~\cite{IceCube}. With these applications in mind, the 
	statistical nature of the photon generating process sets the following natural upper limit 
	on the required amplitude resolution of the readout system. When $N$ is 
	the number of detected photoelectrons, the true number of photoelectrons is in the 
	range $N \pm \sqrt{N}$, i.e.\ the intrinsic relative uncertainty of the detected signal 
	is ${1}/{\sqrt{N}}$. From this it follows that a zero order requirement on the readout system is then to allow the 
	reconstruction of the amplitude/charge of the recorded signal with an uncertainty that is better 
	than ${1}/{\sqrt{N}}$, which we adopt as a benchmark  that FIPSER has to meet or outperform. For the dynamic range we chose 1,000, 
	where aforementioned uncertainty requirement sets the units of
        amplitude to photoelectrons (pe). A dynamic range of 1,000 is
        a typical requirement, and should allow for easy extrapolation
        to applications with different requirements.  In particular, it should be noted that 1,000 is not an 
	upper boundary of the achievable dynamic range. 
	
	While the requirement on the resolution of the reconstructed number of photoelectrons 
	applies to the majority of experiments in astroparticle physics, an equally general 
	requirement on the reconstruction of time does not exist. Therefore, we refrain from 
	formulating such a requirement and, instead, present time related quantities relative 
	to the full width half maximum (FWHM) of the signal to be recorded, i.e. the width of 
	the signal sets the unit of time. This allows for a quick comparison of the simulated 
	FIPSER performance with individual experimental requirements without having to redo 
	the simulations for specific signal widths. 
	
	In the simulations presented here, we use a lognormal distribution to mimic a 
	pulse shape that is typical for a photomultiplier tube (PMT)~\cite{lognorm}.

	\begin{equation}
	\label{eqn:LogNormal}
	f(t)=A\cdot \exp\left(-\frac{[\ln (\omega (t+t_0))]^2}{2\sigma^2}\right)
	\end{equation}

	We will use the FWHM as time unit. With this choice, the 
	sampling rate of FIPSER is given in units of one over the FWHM. For example, a sampling
	rate of two means that FIPSER samples two times per FWHM. 
	
	For the configuration of FIPSER we define $n$ equally logarithmically spaced comparator levels.
	The first level is set at 0.25, and the highest is at 500. The
        $m$-th threshold level is given by:
		\begin{equation}
		\label{eqn:LogSpacing}
		m = 10^{\log_{10}(0.25) + m/(n-1)(\log_{10}(500)-\log_{10}(0.25))}
		\end{equation}

	Our choice of thresholds is arbitrary, and we have not attempted to optimize 
	threshold levels to improve performance. Choosing a logarithmic amplifier coupled to an FADC with $n$ equally spaced comparator levels will yield the same performance. A total of four different configurations of FIPSER are tested, 
	with $n = 8, 12, 16, \text{and} \; 20$.

        We have investigated these configurations in a single pulse and in a two pulse scenario. In the single pulse case, 8,000 lognormal pulses had been generated with a random signal
        amplitude, $A$, between 0.1 and 1,000. Various levels
        of white noise was added to the signal by sampling from a normal distribution centered
        at 0 and varying widths $\sigma$  between 0 and 0.8 in units of
        amplitude. This range of noise levels is representative for experiments like CTA, HAWC, IceCube, MAGIC, VERITAS, which we participated in.
	
\begin{figure*}[ht]
					\centering
					\includegraphics[width=\textwidth]{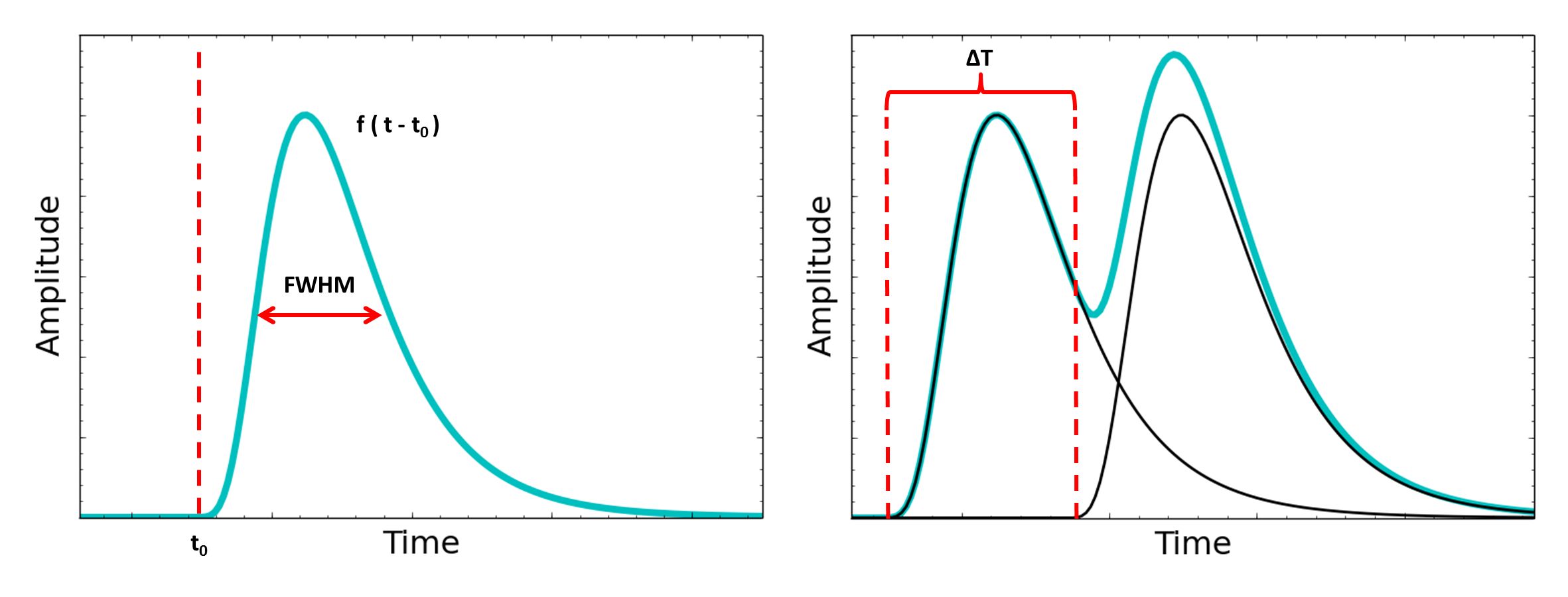}
					\caption{Example traces of the two scenarios that had been simulated to test the FIPSER
                                          concept. The left panel
                                          shows an example of a single pulse being simulated without noise added to the trace. The right panel shows the case of two pulses with equal amplitude (thin lines) being simulated. 
					Their composite waveform is shown in bold as well as the pulse
                                        separation ($\Delta
                                        T$). The trace is shown before noise has been added to the trace.
                                        Some of the benchmarked
                                        parameters are highlighted: 
                                          pulse start time(s), pulse
                                          separation (for two
                                          consecutive pulses), 
}
					\label{fig:pulse_diagrams}
				\end{figure*}

	In the two pulse case, we studied 2,000 sets, each set contains two
        lognormal signal pulses. Both pulses were set to have  equal
        amplitude, while the separation between the two was chosen
        randomly between 0.25 and 2 FWHM. White noise was added to the signal
	with $\sigma = 0.1$. Figure~\ref{fig:pulse_diagrams} shows example traces of the
        two types of simulations we have conducted. 



The simulated traces are sampled at a given FIPSER sampling rate and discretized based on the FIPSER threshold levels.
Both leading and trailing edges of the pulse are identified
as such. 


        The simulated data had been reconstructed with two independent techniques. 
	One technique uses the least-square method with the known signal shape 
	as a fit template. The second technique is a probabilistic method, essentially a maximum 
	likelihood method, that uses the additional information about the threshold 
	settings of FIPSER and the characteristics of the noise when reconstructing the signal. 
	Performance was gauged based on how accurately the pulse time
        and amplitude could be reconstructed. Both methods yield
        comparable performance  
	in regards to these metrics. Presently, only the least square method allows for the examination 
	of two consecutive pulses to determine their amplitude and separation. A detailed description 
	of the reconstruction methods is given in the following two sections. 

\section{Least Square Reconstruction Method}
\label{sec:square}

	In the least square method the pulse time and amplitude are reconstructed with a 
	$\chi^{2}$ inspired test that compares FIPSER data sets to the known pulse shape with different amplitudes and times. 

	\subsection{Single Pulse Reconstruction}
\label{subsec:square_single} 
		
		In the event that a simulated pulse crosses only one comparator, 
                an exact solution exists. An explicit solution uses less computational time than a 
		minimization. The minimum signal amplitude that can be recorded with FIPSER is obviously determined by the comparator with the lowest threshold.


                In the case when only one comparator fires the duration during which the signal amplitude is above the comparator threshold (time over threshold, ToT) is used to find the pulse amplitude. This is done by scaling the template pulse until a match in 
ToT is obtained. After the ToT has been matched the time and the pulse amplitude of the scaled template pulse provide the reconstructed time and amplitude. 
		
		If more than one comparator fires a full fit is performed. In the first step of the fit procedure the limits of the fit parameters are set.	
		The pulse arrival time is constrained to within $-{5}/{4}$ FWHM and ${1}/{4}$ FWHM of the time the first 
		comparator fires. The limits are based on the fast rising edge of the simulated lognormal signal, 
		and had been fine tuned by hand. The reconstructed amplitude is constrained to within the highest 
		comparator which triggered, and the comparator level directly above, with 1,000 being used as a maximum
		in the event that all comparators have triggered. 
		
		After establishing the parameter limits, a coarse
                uniformly sample scan
                of the parameter space is performed to determine the
                initial condition for the minimization. In this step,
                28 possible pulses are compared to the FIPSER
                data. The 28 pulses are combinations of four time
                values and seven amplitude values that fall within the determined parameter limits. 
		
		For FIPSER, with limited number of comparators, it is
                plausible for the $\chi^2$ to be exactly zero. In the
                case when the coarse scan returns an already 
                perfect match to the comparator crossings the next
                steps are skipped and the results from the coarse scan
                returned as the final result.  
		
		To assess the goodness of the match between the test pulse and the FIPSER data we evaluate 
		\begin{equation}
		\label{eqn:ModChi}
		\sum\limits_{levels} \bigg [ \bigg ( \sum\limits_{crossings}\frac{(t_{test} - t_{true})^{2}}{\sigma_{1}} \bigg ) \; \textrm{or} \; \bigg ( \frac{(ToT_{test} - ToT_{true})^2}{\sigma_{2}} + C \bigg ) \bigg]
		\end{equation}
		
		Here, $t$ is the time when a comparator crossing is registered by FIPSER, $ToT$ is the total time over
		threshold, and $\sigma$ is a constant. $C$ is also a constant, and is chosen 
		such that it exceeds the largest expected value of the leftmost sub-equation.
		
		Equation~\ref{eqn:ModChi} was conceived to compare pulses which have a different
		number of comparator crossings. An important aspect of this equation is that it 
		assumes each upward comparator crossing has an associated downward crossing, i.e. 
		there are no missing trailing edges. 
		
		To evaluate Equation~\ref{eqn:ModChi} on a single
                comparator level, we first determine 
		if the level is crossed the same number of times by both the true pulse and the test 
		pulse. If it is crossed the same number of times, the leftmost sub-equation is used. 
		Here, $t_{test}$ and $t_{true}$ are the times of corresponding comparator crossings. 
		To determine which crossings are to be considered
                corresponding, we break the level data 
		into two lists containing upward and downward crossings. The earliest upward crossing in 
		the test data set corresponds to the earliest upward crossing in the true data set, and 
		so forth. Performing this over all points and summing produces a numerical value for 
		that level. 
		
		In the event that the level is not crossed the same number of times by the test and true 
		pulses, we use the rightmost sub-equation. Each level will have only one data point, representing
		the total ToT. The addition of the constant $C$ assures that the numerical value for this
		sub-equation is always larger, and thus less preferable, than in the case where the comparator
		level is crossed the same number of times. It is important to note that the ToT only relays
		information about pulse	amplitude,	and not the time that the pulse occurs. As such, this is
		considered an intermediate step. When the test and true pulses are sufficiently similar, each
		comparator will have the same number of crossings, and the preferable leftmost sub equation can
		be used for the level. The final step of a succesful
                reconstruction is a $\chi^2$ test using the left side
                term of Equation \ref{eqn:ModChi}.
		
		Once this procedure has been followed for all comparator levels, the individual values
		are summed. 
		
		Comparing the time over threshold 
		allows Equation~\ref{eqn:ModChi} to be easily applied to multiple pulse waveforms. 
		For example, while a single pulse waveform can cross a comparator level either 
		twice or not at all (disregarding noise), a waveform composed of two pulses can 
		cross a comparator level twice, four times, or not at all.  
		
		Once the parameter limits and the first guess have been determined, 
		pyMinuit~\cite{pyminuit} is used to minimize Equation~\ref{eqn:ModChi}.

			\begin{figure*}[!ht]
				\centering
				\includegraphics[width=\textwidth]{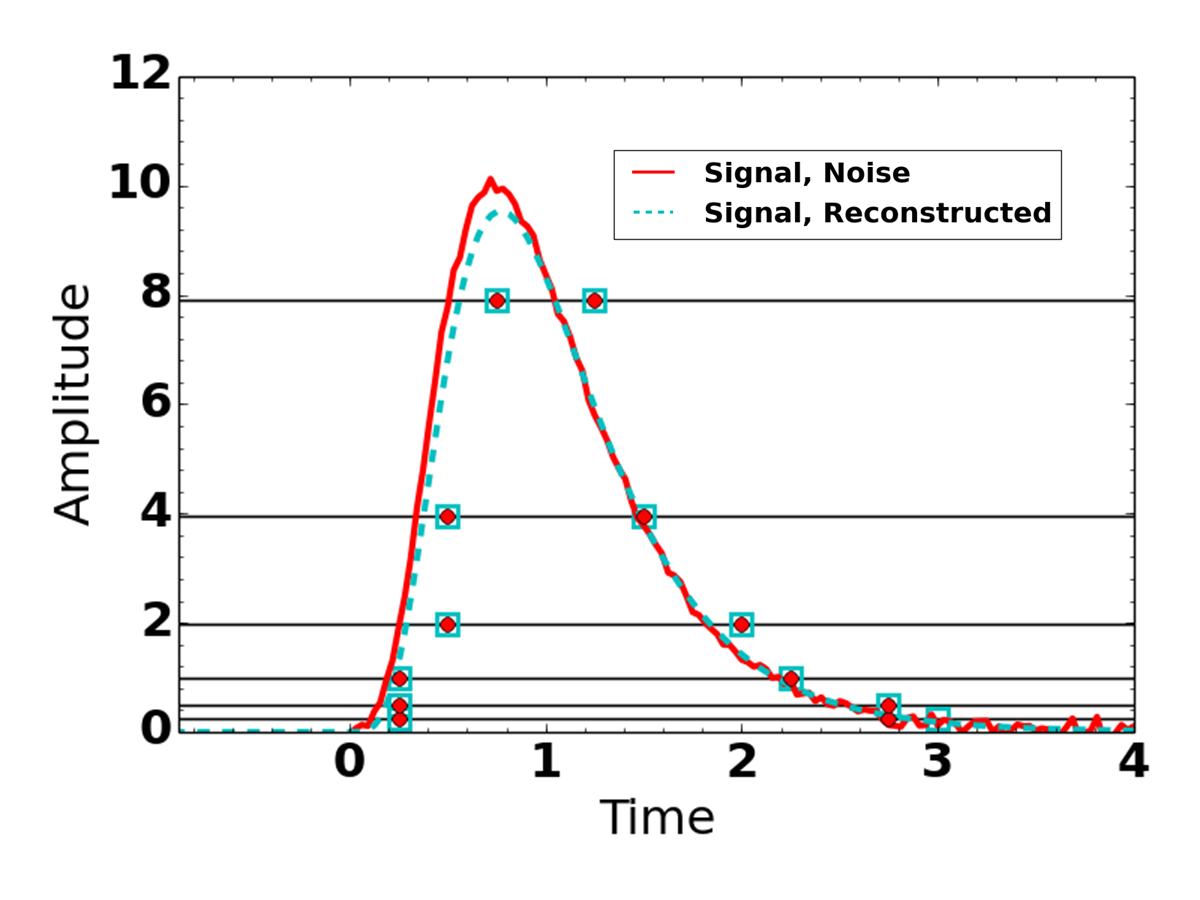}
				\caption{A reconstructed pulse (dashed) and original pulse (solid), with a noise level of $\sigma = 0.1$. 
						The circles show when FIPSER would have recorded the comparator level crossing for the original pulse, which samples the comparator outputs with a rate of 4 per FWHM. The comparator levels are shown as horizontal lines.
						Squares show when FIPSER would have recorded the comparators logic high states for reconstructed pulse.}
				\label{fig:single_fitting}
			\end{figure*} 
	
		An example of a pulse reconstructed with the least square method is shown in 
		Figure~\ref{fig:single_fitting}. The solid curve shows the signal with noise that 
		is digitized with FIPSER (circles, placed at the end of the sampling period). 
		The least square method best fit is shown with the sampled data 
		points represented as circles. The best fitting pulse is shown as a dashed line and
                the matching FIPSER threshold crossing times are shown
                as squares. The best fit pulse is a
                good match to the original pulse shape. The differences can be attributed to fluctuations caused by noise.
		
		While this reconstruction method works well with a noise level of $\sigma = 0.1$, 
		a fraction of the data cannot be successfully be
                reconstructed at higher noise levels. The failure
                rates are modest, so we have developed criteria to
                remove failed reconstructions. A reconstruction
                is considered failed with the following characteristics:
                an amplitude greater than 60 units, an 
		amplitude less than 2 amplitude units below a comparator or 1000 amplitude units, and a 
		modified $\chi^{2}$ above a predetermined value. Only
                a single reconstruction out of 8000 failed for noise level of
                $\sigma = 0.1$ or lower. With noise levels of $\sigma = 0.2, 0.4, 0.6, \text{ and } 0.8$, 
		0.38\%, 2.68\%, 4.72\%, and 5.40\% of 8000 reconstructions
                fail. We do not apply this criteria for double pulse
                reconstructions. All single pulse figures presented
                here exclude data for failed reconstructions.
		
		\subsection{Two Pulse Operation}
\label{subsec:square_double}
		
			The procedure for analyzing two pulses is nearly identical to that of reconstructing a single pulse, 
			with the only differences being how the parameter limits and the initial guess is being determined.
			The procedure requires the user to specify that two pulses should be reconstructed, and does not currently contain 
			the logic to autonomously determine the number of pulses to be fit. 
		
			The arrival time of the first pulse is constrained to be between  $-{5}/{4}$ and ${1}/{4}$ FWHM 
			of the time when FIPSER registers that the first comparator has triggered. 
                        The time of the second pulse is constrained 
			to be between the time the first pulse triggers and the time the signal drops again below the lowest 
			threshold. But has a maximum of 5 FWHM separation. The amplitude of both pulses 
			is constrained to be between 0.25 and the comparator with the next highest level that has not triggered, 
			or 1,000 if all comparators have triggered. It is not a priori assumed that both pulses 
			have the same amplitude. 
			
			The number of test pulses in the coarse scan is allowed to vary. The 
			arrival time of the first pulse is fixed to be half of $1/\textrm{Sampling Rate}$
			before the first comparator triggered. The arrival time of the second pulse is varied with 
			a minimum spacing of ${1}/{8}$ FWHM between the first and last comparator triggered by a charging signal, 
			with a maximum scan range of 5 FWHM. Both pulse amplitudes are scanned uniformly with 7 different 
			values between 0 and their maximum boundary condition, and are not assumed to be the same.

\section{Probabilistic Method}
\label{sec:prob}

	In our second method to extract signal time and amplitude, we maximize the
	 probability that the discretized values can be described with a known signal 
	 shape $f(t)$, an unknown time $t_0$, and amplitude $A$. The method takes 
	 uncorrelated and white noise into account with a known amplitude distribution $\nu(x)$.
	
	Here is an example how to calculate the probability that a discretized pulse shape 
	is due to a pulse with amplitude $\alpha$ and time $\tau_0$. Lets consider the 
	time $t_0$ when a discriminator fires for the first time. For that time the probability 
	$P$ to observe a noise amplitude $X_0$ can be expressed as 
	
	\begin{equation}\label{1}
	 P(X_0|\alpha,\tau)=\int_{L_0-\alpha f(\tau_0 )}^{U_0-\alpha f(\tau_0)} \nu(x_0)\ dx_0,
	\end{equation}
	
	where $L_0$ is the level of the discriminator with the highest threshold level that fired at time $t_0$ 
	and $U_0$ is the next highest threshold of the discriminator that did not fire. The 
	above equation only considers the situation at time $t_0$ and all the following time 
	samples have also to be included in order to calculate a combined probability. 
	$P(X_i)$ is drawn from the noise distribution of amplitudes and it is assumed 
	that the noise between FIPSER samples is uncorrelated. It thus follows that the combined probability $Q$ for 
	an assumed $\alpha$ and $\tau$ is the product of all the $P(X_i)$,\emph{i.e.}
	
	\begin{equation} \label{2}
		Q(\alpha,\tau_0)=\prod\limits_{k=0}^{n-1} \int_{L_k(\alpha ,\tau_0)}^{U_k(\alpha ,\tau_0)} \nu(x_k)\ dx_k,
	\end{equation}		
	
	where $k$ runs over all the $n$ FIPSER samples during which at least one discriminator fired.
	
	To properly normalize the probability Q($\alpha,\tau_0$) we need to integrate $Q$ over all possible amplitudes 
	$A$ and times $t$. May the allowed range for $t$ over which $Q$ is integrated be between $t_0^m$ and $t_0^M$ and the allowed range for the signal amplitudes be between $A_m$ and $A_M$. Guidance on how to chose values for these limits is given 
	later. The normalization is then
	\begin{equation}\label{3}	
	N\equiv \int_{A_m}^{A_M}\int_{t_0^m}^{t_0^M} Q(\alpha,\tau_0)d\tau_0 d\alpha\,.
	\end{equation}
	In order to arrive at the amplitude and time of the discretized signal we chose to use the expected values
	\begin{equation}
	\langle t_0 \rangle=\frac{1}{N}\int_{A_m}^{A_M}\int_{ t_0^m}^{t_0^M} \tau_0Q(\alpha,\tau_0)d\tau_0 d\alpha
	\end{equation}
	and
	\begin{equation}		
	\langle A \rangle=\frac{1}{N}\int_{A_m}^{A_M}\int_{t_0^m}^{t_0^M} \alpha Q(\alpha,\tau_0)d\tau_0 d\alpha
	\end{equation}
		
	\subsection{Practical Considerations}
\label{subsec:pract}

	The description of the fit method is complete and its performance depends on the ability to calculate the 
	above integrals fast and with high enough precision. In the following we share some of our experience optimizing
	precision and computation time.
		
	\paragraph{Optimizing integration}
	When the sample period $T$ is sufficiently small and the number of thresholds $N$ large 
	enough, the integral in (\ref{2}) is close to zero whenever $\tau_0$ and $\alpha $ are not very 
	close to a reasonable initial time and amplitude pair. This is due to the fact that faster sampling
        and more thresholds yield more precise reconstructed pulses, and thus the 
	vast majority of the allowed parameter space $[t_0^m,t_0^M]\times [A_m,A_M]$ contribute little 
	to the total integral. In other words, the biggest contributors to the integral are a small number 
	of time and amplitude pairs. For this reason, standard quadrature methods, 
	unless very heavily subdivided, perform poorly. This may be remedied by reducing the 
	integration space to one which includes a much larger density of reasonable amplitudes and times.
		
	One method to find \emph{dense} regions in the parameter space is to firstly maximize equation \ref{2} as a 
	function of $\tau_0$ and $\alpha $. Then, the user may integrate a small area around that maximum. However, 
	caution needs to be exercised because the parameter space might have several local maxima, which can cause the 
	maximization method to fail finding the global maximum.

	\subsection{Implementations for Log-Normal signal shape}
	The log-normal signal as parametrized by the initial time and amplitude is given by 
	Equation~\ref{eqn:LogNormal}. For this pulse shape we chose limits on the initial time 
	of $t_0^m=-2$, $t_0^M=1$. The motivation for setting the lower limit at minus two times 
	the FWHM of the signal instead of choosing 0 is that it allows for sufficient time to 
	consider whether discriminators firing due to noise are part of the signal or not. Setting 
	$\omega ={8}/{7},f(t)$ reaches its maximum when $t={1}/{w}={7}/{8}$, and thus 
	$t_0^M=1$ is slightly greater than the peak time and for the vast majority of signal arrivals 
	at least one discriminator will have fired by $t=t_0^M+T$.\\
			
	The amplitude limits will simply be $A_m=0.1$ and $A_M=1000$ in this study. In principal 
	this range can be narrowed for each case to be roughly in between the highest discriminator 
	level that fired and the next highest that did not fire. The exact range has to be larger 
	than that because the discriminator is fired by the sum of the signal and noise.
	
	For the probability distribution of noise we assume a Normal distribution
	\begin{equation}\label{10}
	v(x)=\frac{1}{\sigma_{\nu }\sqrt{2\pi }}\exp\left(-\frac{x^2}{2\sigma_\nu ^2}\right)
	\end{equation}
	with $\sigma_\nu $=0.1. 
	
	In our simulation studies we use an iterated Gauss-Kronrod numerical integration method 
	with 20 linear steps in time and 30 steps with respect to the square root of the amplitude  
	achieved a reconstructed amplitude and initial time pair.

\section{Results}	
\label{sec:results}

	We found that the probabilistic method performs slightly better than the least square
        method in the reconstruction of single pulses. However, the probabilistic method was not tested on the 
        reconstruction of the two pulse scenario, which is why we only show results produced with 
        the least square methods to be consistent throughout the paper.
	
	\subsection{Single Pulse Charge Resolution}
\label{subsec:sCharge}
	
\begin{figure*}[ht]
					\centering
					\includegraphics[width=\textwidth]{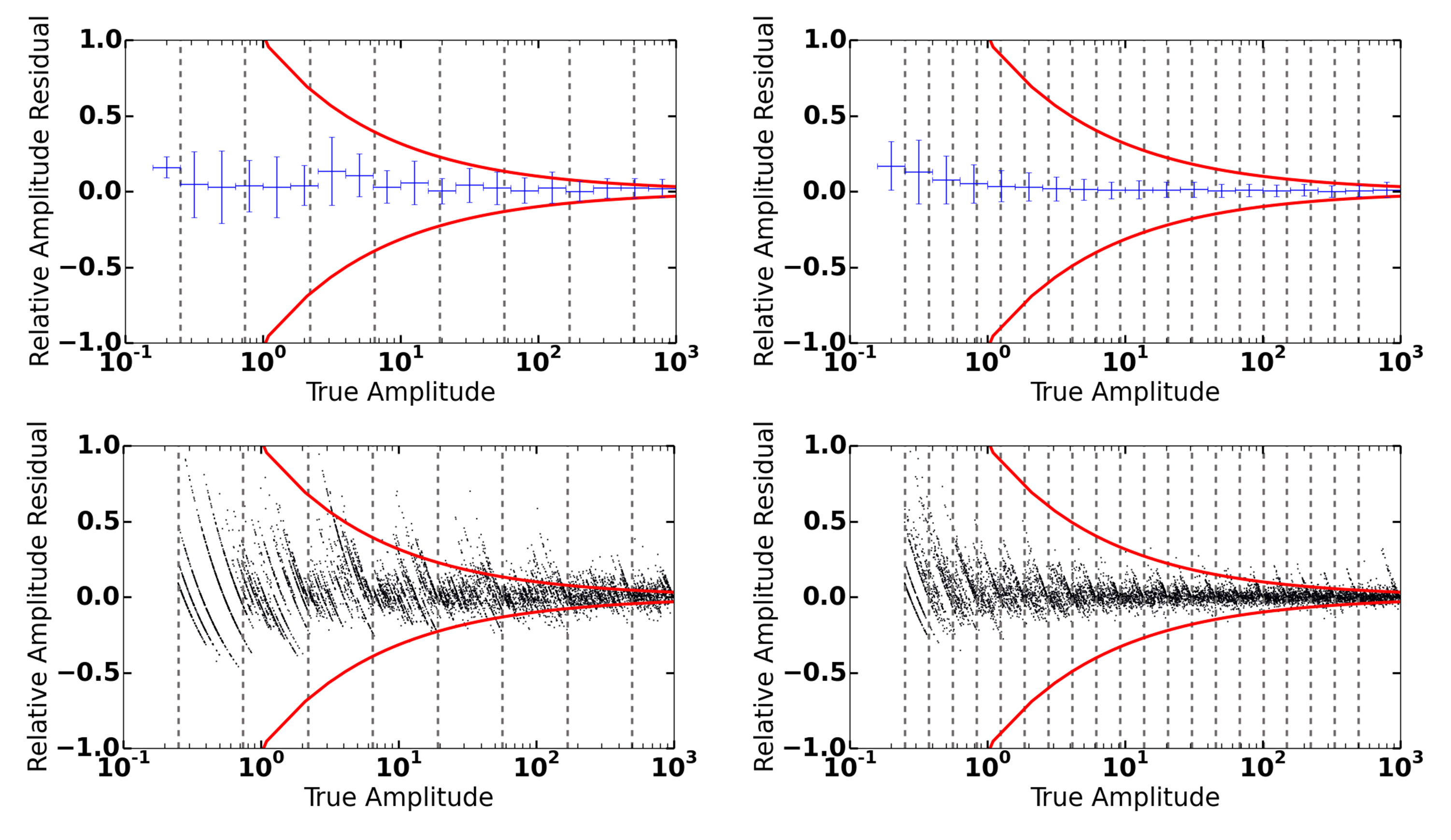}
					\caption{The relative amplitude residuals resulting from reconstructing simulated pulses with a FIPSER configuration that uses 8 comparators (left) and 20 comparators (right), respectively. 
							The simulated noise level was 
							$\sigma=0.1$, and a sampling rate of 4 per FWHM was used.  The bottom panels 
							show scatter plots of the residuals, and the top panels show the same data binned in true amplitude and showing the mean and root mean square of all data points in each bin (profile histograms). The vertical dashed lines show the comparator levels, and the bold curve marks the Poisson limit. }
					\label{fig:8_20_amp}
				\end{figure*}

		The accuracy of reconstructing the signal amplitude is judged by evaluating the relative
		amplitude residuals $(A_{rec}-A_{true})/A_{true}$ between the reconstructed amplitude $A_{rec}$ 
		and the true amplitude of the signal $A_{true}$. Figure~\ref{fig:8_20_amp} shows two 
		example scatter plots of the residuals as a function of $A_{true}$. Results for a FIPSER configuration 
		with 8 comparators is shown on the left, and for a configuration with 20 comparators on 
		the right. The Poisson limit, as defined in Section~\ref{sec:config}, is indicated by 
		curved dashed lines, and is satisfied if $68\%$ of all events fall within these bounds. An 
		analysis of these specific results shows that a FIPSER configuration with 8 comparators does not meet
		the Poisson limit requirement for pulse amplitudes greater than 100, while pulse amplitudes 
		up to 600 can be reconstructed within the Poisson limit fall if 20 comparators are used. Both simulations 
		had been performed with a sampling rate of 4 samples per FWHM and with white noise of 
		$\sigma=0.1$, which is typical for noise that is introduced in the signal chain and is not due to detector signals. 
		
\begin{figure*}[ht]
					\centering
					\includegraphics[width=\textwidth]{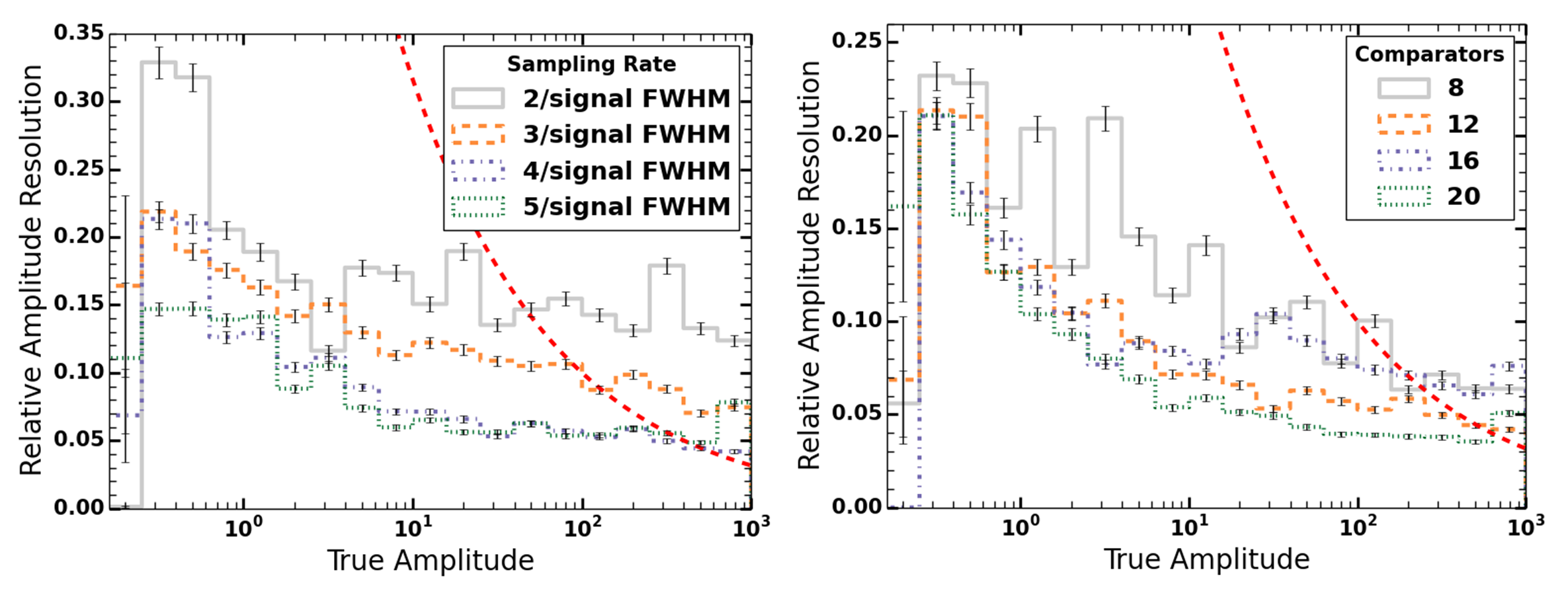}
					\caption{Amplitude resolution for a scenario with a single pulse in the trace and
					white noise with $\sigma = 0.1$. In the left panel the FIPSER configuration uses 12
					comparators with varying sampling rates, and in the right panel the sampling rate is
					fixed at 4 per FWHM. The curved dashed line shows the Poisson limit.}
					\label{fig:single_amp}
				\end{figure*}

		To allow for a direct comparison between different FIPSER configurations and 
		the Poisson limit, the scatter plots are re-binned in true amplitude, and 
		the mean and root mean square (RMS) of the residuals are calculated for each bin. 
		The right panel in Figure~\ref{fig:single_amp} shows the RMS values vs. true amplitude for 
		configurations with 8, 12, 16, and 20 comparators, which includes the configuration already shown  
		in Figure~\ref{fig:8_20_amp}. All of these simulations had been performed with a 
		sampling rate of 4 samples per FWHM and a white noise level of $\sigma=0.1$. 
		As expected, the amplitude resolution improves with more comparators. 
		For the configurations with 12 and 20 comparators, the amplitude
                resolution meets the Poisson limit requirement up to an
                amplitude of $\sim$500.

		We note that the configuration with 16 comparators produces worse results 
		than with 12 comparators at amplitudes greater than 10. This feature is present in both 
		the least square and probabilistic reconstruction methods. 
        We believe this is a result of a less-than-ideal spacing of the comparator levels, and
        hypothesize that distributing 
		the levels in a different way would remedy this. However, we chose to 
		leave this as an avenue for a future study. 
		
		A dominant feature visible in Figure~\ref{fig:8_20_amp} is the banded structure in 
		the scatter plots. This is due to the finite sampling
        rates, which cause pulses in specific amplitude ranges to of have identical thresholds
        crossing times in FIPSER. 
		As such, the minimizer used in the Least Square method only returns a set of discrete 
		reconstructed amplitudes. The probabilistic reconstruction method is free of these artifacts. 
		
		The left panel of Figure~\ref{fig:single_amp} shows how charge resolution 
		varies with sampling speed. The resolution does not improve with sampling 
		rates higher than 4 samples per FWHM. Figure~\ref{fig:8_20_amp} and 
		Figure~\ref{fig:single_amp}  combined show that it is more efficient to 
		improve the amplitude resolution of FIPSER by increasing the number	of comparators, 
		as opposed to increasing the sampling rate. 
			
			\begin{figure*}[ht]
				\centering
				\includegraphics[width=\textwidth]{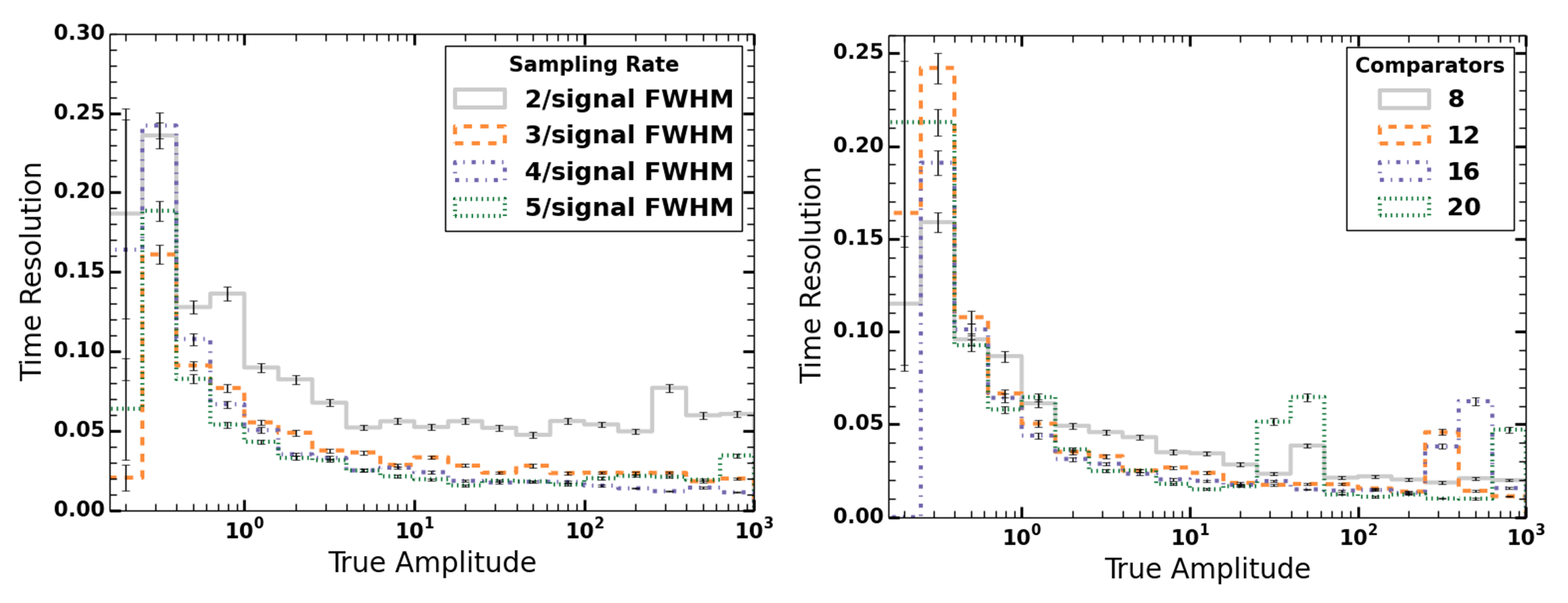}
				\caption{Time resolution for a scenario with a single pulse in the trace and
						white noise with $\sigma = 0.1$. In the left panel the FIPSER configuration
						uses 12 comparators with varying sampling rates, and in the right panel the
						sampling rate is fixed at 4 per FWHM.} 
				\label{fig:single_time}
			\end{figure*}

\begin{figure*}[ht]
				\centering
				\includegraphics[width=\textwidth]{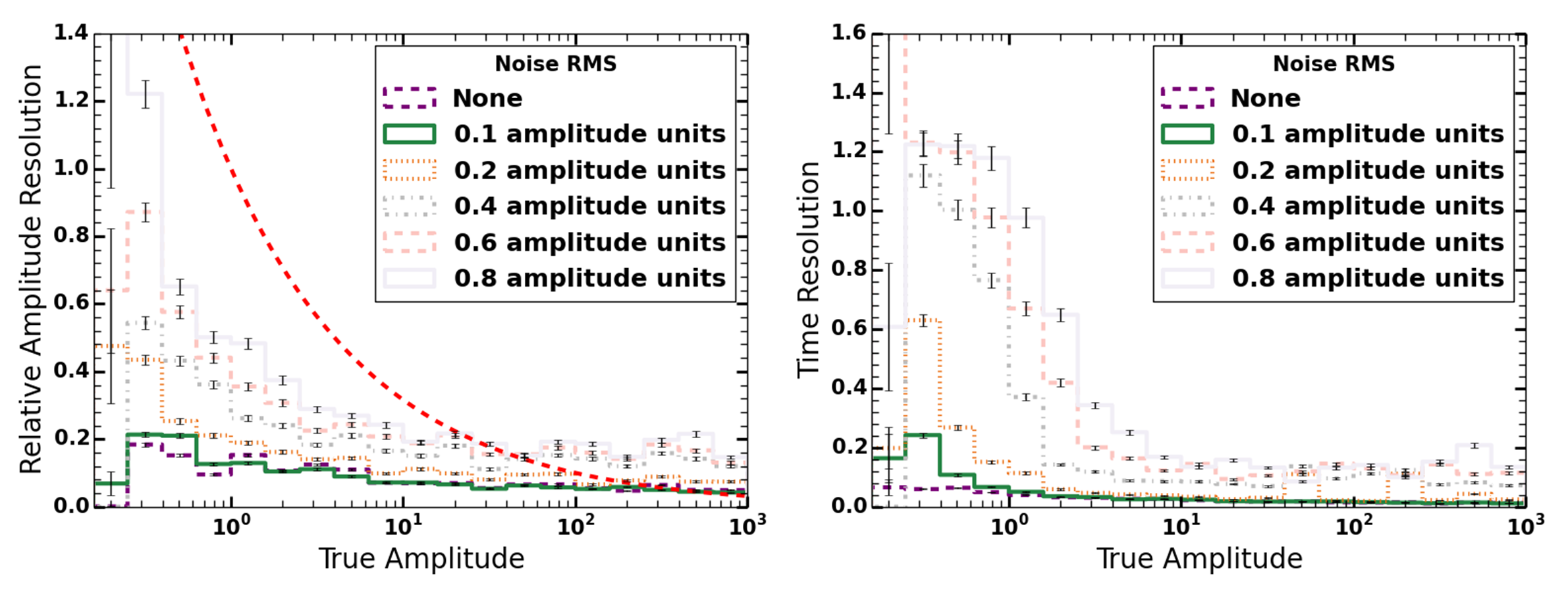}
				\caption{Results for a FIPSER configuration with a sampling rate of 4 samples per
                                 FWHM and 12 comparators, while simulating five different noise levels.
                                 The curved dashed line in the left panel 
				shows the Poisson limit. }
				\label{fig:noise}
			\end{figure*}

		Figure~\ref{fig:noise} shows the performance of FIPSER at varying noise levels while 
		utilizing 12 comparators at a sampling rate of 4 samples per FWHM. The RMS of the noise is 
		varied between 0 and 0.8 amplitude units, with the latter being a worst case scenario for most 
		applications in particle astrophysics. The Poisson limit requirement for the amplitude
		resolution is met for most of the dynamic range. Only for signal amplitudes above 40 is 
		the amplitude resolution worse than the Poisson limit when noise levels are higher than 0.4~$\sigma$.

	\subsection{Single Pulse Time Resolution}
\label{subsec:sTime}

\begin{figure*}[ht]
				\centering
				\includegraphics[width=\textwidth]{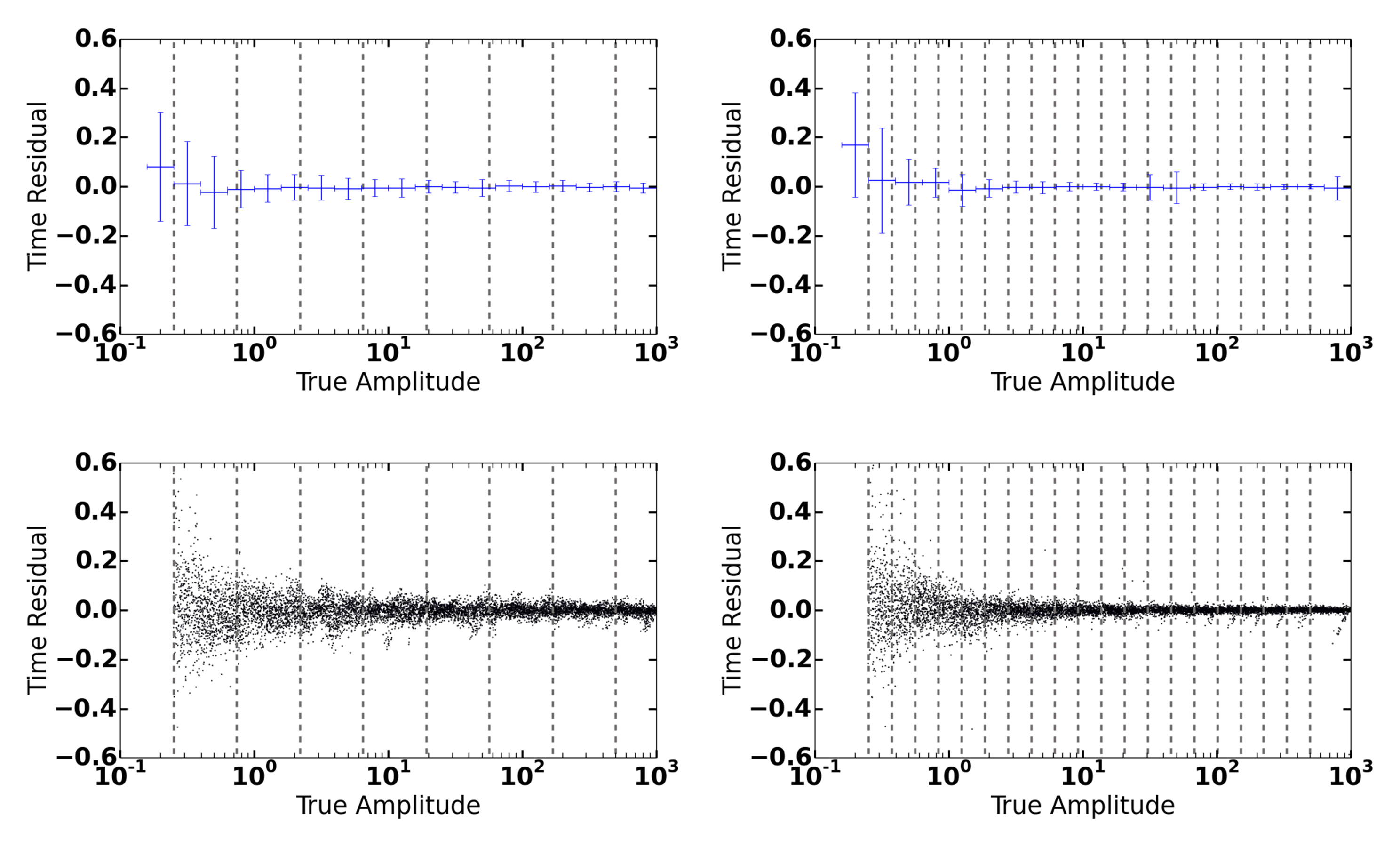}
				\caption{The time residuals for a configuration with 8 comparators
						(left) and 20 comparators (right). The top row panels
						show the mean and the standard deviation of the time residuals in bins of
                                                true signal amplitude. The bottom panels show scatter plots 
						of the time residuals.  The comparator levels are indicated by the vertical dashed lines.}
				\label{fig:8_20_time}
			\end{figure*}

		The ability of FIPSER to accurately determine the timing of a pulse is qualitatively 
		assessed by calculating the difference between the true time and the reconstructed 
		time. 
                Figure~\ref{fig:8_20_time} shows scatter 
		plots of the time residuals for the same FIPSER configurations as in Figure~\ref{fig:8_20_amp}.  
		As before, we divide the scatter plots into bins of true amplitude and calculate the 
		mean and root mean square (RMS) of the time differences. These results, 
		along with two additional FIPSER configurations, are shown in the right 
		panel of Figure~\ref{fig:single_time}. 
		It can be seen in the right panel of Figure \ref{fig:8_20_amp} 
        that 12 comparators achieve a significant improvement in time resolution than using a
        configuration with 8 comparators, particularly above amplitudes of 1. 
		
		The time resolution improves dramatically from 0.03 FWHM for a configuration with 8 comparators 
		to 0.01 FWHM for a configuration with 20 comparators for large signal amplitudes. We note that 
		for a detector signal with a FWHM of 10\,ns, these values translate into time resolutions of
		300\,ps and 100\,ps, respectively. The sampling rate of 4 per FWHM equates 
		to a moderate 400 Megasamples per second. For comparison, PMT signals in IceCube are 
		digitized at 333 Megasamples per second. 

		\begin{figure*}[ht]
			\centering
			\includegraphics[width=\textwidth]{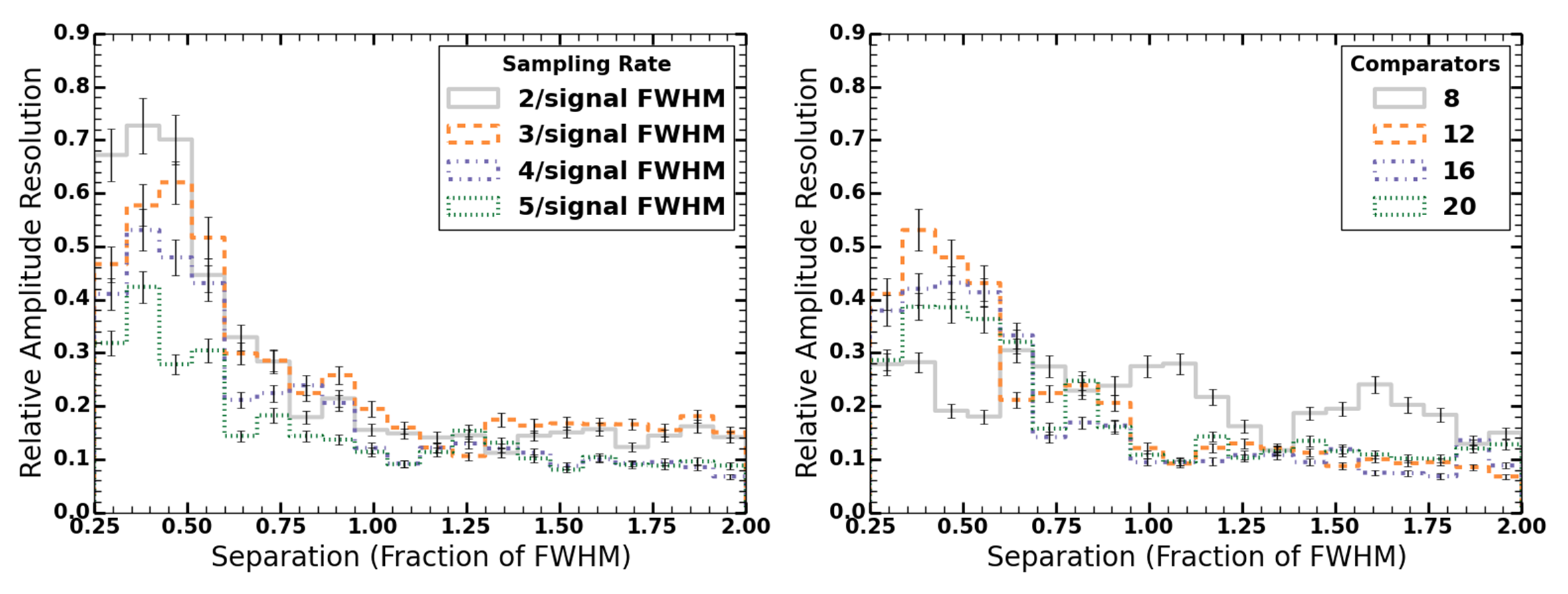}
			\caption{Amplitude resolution of the 
			first pulse for a scenario with two pulses in the trace and white noise with $\sigma=0.1$. 
                        In the left panel the FIPSER configuration uses 12 comparators with varying sampling rates and in the right panel the
                        sampling rate is fixed at 4 per FWHM while the number of comparators is varying.}
			\label{fig:fa}
		\end{figure*}		
				
		The left panel of Figure~\ref{fig:single_time} shows how the time resolution improves with
		increasing sampling rate for a configuration with 12 comparators.  At larger amplitudes the resolution remains relatively constant. 
		
		The same conclusion we found when studying the amplitude resolution also applies to the time resolution, it seems more practical to choose a 
		configuration with more comparators and a moderate sampling rate instead of increasing 
		the sampling rate while keeping a smaller number of comparators.

	\subsection{Double Pulse Charge Resolution}
\label{subsec:dCharge}

		The performance of FIPSER for the reconstruction of two pulses per trace can be evaluated
		by looking at the amplitude resolution of the first pulse, the second pulse, and the 
		aggregate waveform. Because only two of them are independent it is sufficient to analyze the 
		amplitude resolution of the first pulse and the aggregate waveform.
                In our studies we found that the performance 
		of reconstructing two pulses is relatively uniform for signals with amplitudes above 
		1, \emph{i.e.} for signals that fire several comparators. We, therefore, fixed the amplitude of both 
		the first and second pulse to 10 and tested the reconstruction performance 
                as a function of the distance between the first and second pulse. 
		

			\begin{figure*}[ht]
				\centering
				\includegraphics[width=\textwidth]{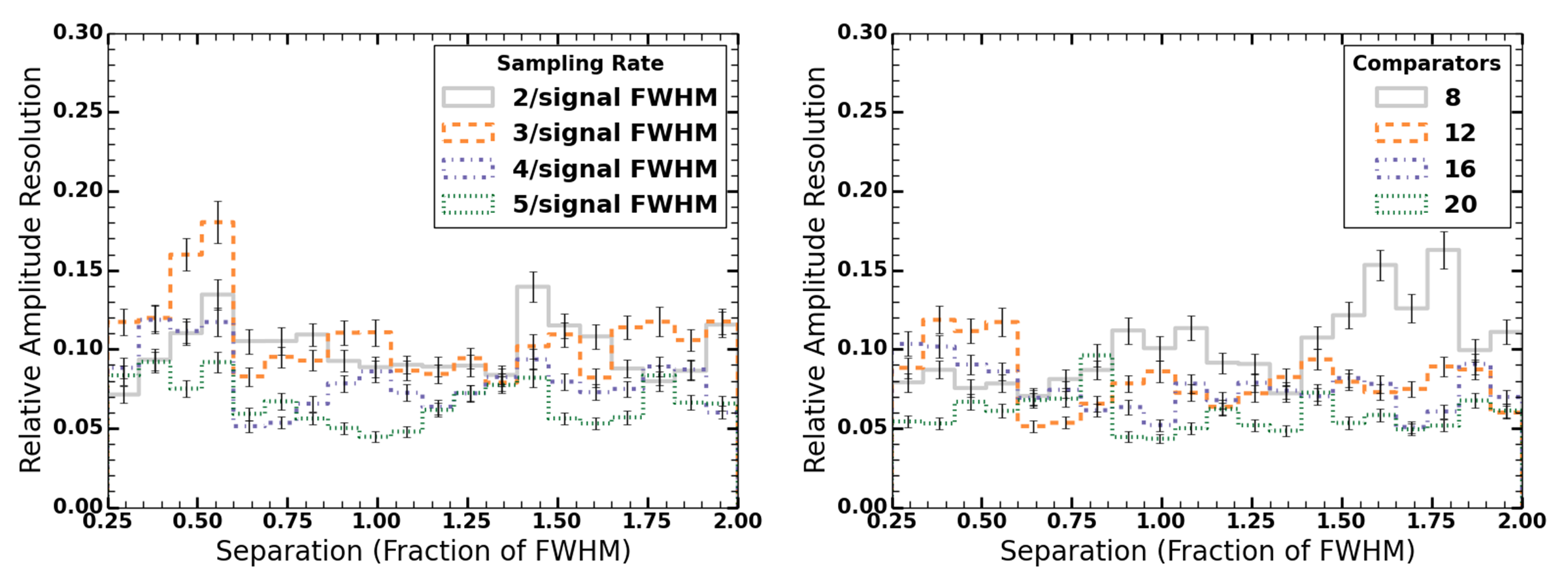}
				\caption{Amplitude resolution for reconstructing the total 
								signal in a trace that contains two pulses with a white noise level of $\sigma=0.1$. 
                                FIPSER is configured with 12 comparators in the left panel with varying sampling rates, whereas
                                the sampling rate is fixed to 4 per FWHM in the right panel and the number of comparators is being varied.}
				\label{fig:tc}
			\end{figure*}
		
		Figure~\ref{fig:fa} shows the amplitude resolution of the first reconstructed pulse as a function of pulse separation.
                The amplitude resolution slightly improves with increasing number of comparators and increasing sampling rates.

        Interestingly it is the configuration with 8 comparators that now appears to be worse than one would expect from the 
        resolutions obtained with the other configurations. If the origin of the worse resolution is the same as in the single pulse case a different spacing of the comparator levels
        might yield a better resolution for the 8 comparator case. In fact, for the given fixed pulse amplitude of 10 
        the FIPSER configurations with 12, 16, and 20 comparators all have a comparator level close to the peak of the combined amplitude of the two peaks, which is not the case in the configuration with 8 comparators.

		Figure~\ref{fig:tc} demonstrates how the charge resolution of the two pulses combined varies for
        different sampling rates and number of comparators. Overall, the resolution is fairly consistent, 
        and does not show any obvious trends over the range of sampling rates, number of comparators, 
        or pulse separations which were tested.

	\subsection{Double Pulse Time Resolution}
\label{subsec:dTime} 
	
			\begin{figure*}[ht]
				\centering
				\includegraphics[width=\textwidth]{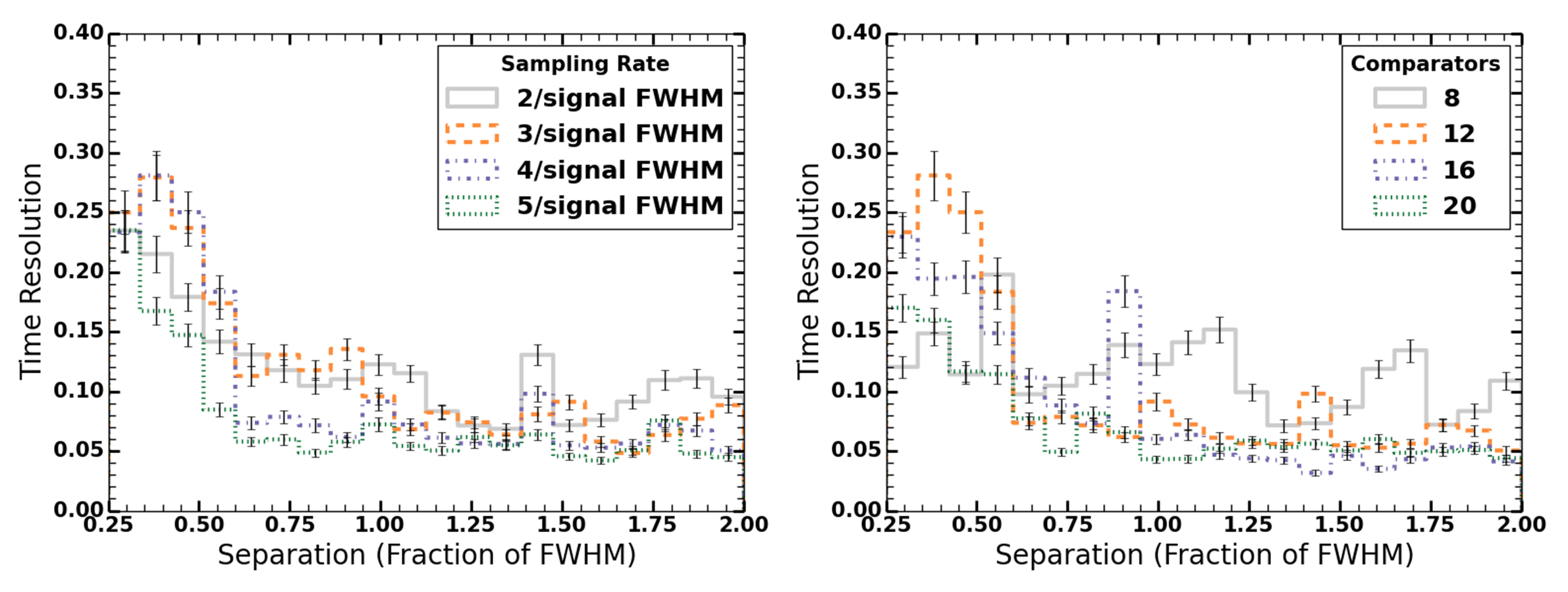}
				\caption{Resolution of reconstructing the separation between two pulses
				 with white noise  ($\sigma=0.1$) added to the traces. 
				 In the left panel FIPSER is configured with 12 comparators and varying sampling rates, whereas 
                                 in the right panel the sampling rate is fixed at 4 per FWHM and the number of comparators is varied.}
				\label{fig:ps}
			\end{figure*}
		
		Particles of an air shower arrive at different times. 
		In particular, muons precede electromagnetic components.
		The ability to resolve consecutive pulses may aid in the detection of muons in detectors such as HAWC. 
		For the analysis of the achievable time resolution in the two pulse case we evaluated the
        resolution of the time of the first pulse and the time resolution of the separation between the two.
        As can be seen in 
		Figure~\ref{fig:ps}, the determination of the pulse separation improves only slightly with pulse separation once 
        the pulse separation is more than $\sim0.5\,$FWHM. Some improvement in resolution is seen when the sampling rate
        increases from 3 to 4 samples per FWHM.
 

\section{Conclusions}
\label{sec:con} 

	We performed conceptual studies of FIPSER, a readout concept, which promises to achieve significant 
        power savings compared to FADC based readout systems. The results of two independent reconstruction methods show that 12 
	comparators at a moderate sampling rate of 4 samples per signal FWHM can meet the 
	$\frac{1}{\sqrt{N}}$ requirements over a dynamic range of three orders of magnitude. 
	A time resolution significantly better than 1\,ns seems possible for pulses with a 
	FWHM of less than 10\,ns. The same conclusions can be drawn when the trace is 
        composed of two partially overlapping pulses.
	
	A limitation of the FIPSER concept is that the pulse shape needs to be known beforehand. 
	While this should not pose a problem for most applications in astroparticle physics, the 
	concept needs to be studied in greater detail for applications in which pulses 
	of similar amplitudes can overlap frequently and the pulse shape cannot be assumed fixed. 
        More sophisticated reconstruction 
	algorithms could mitigate some of these limitations. 
	
	Compared to established readout schemes, FIPSER provides a number of practical 
	advantages. Due to a decrease in the number of comparators by an order of 
	magnitude, FIPSER has the potential to realize significant power savings when 
	compared to existing readout systems. Other positive features are compactness of a 
	FIPSER readout and a possible reduction in data volumes.  
	FIPSER is dead time free, and it is straightforward to implement online event 
	selection and processing. 
	
        The implementation of a prototype of FIPSER is beyond the scope of this paper. 
        A possibility for implementing the concept is to use FPGAs \cite{gary5}, which have developed into one of the most versatile tools for data acquisition systems in recent years.



\section*{Acknowledgements}
We acknowledge support by Georgia Tech's GT-Fire program.


\begin{thebibliography}{1}
\bibitem[Analog Devices (2016)]{AD}
Analog devices (2016)
{\em www.analog.com}
\bibitem[Tibaldo et al.\ (2015)]{gary1}
L.~Tibaldo, and et~al.
{\em 34th International Cosmic Ray Conference (ICRC)}, (2015)
\bibitem[Bechtol et al.\ (2012)]{gary2}
K.~Bechtol, and et~al.
{\em Astroparticle Physics}, \textbf{36}, 156--165 (2012)
\bibitem[Brenton et al.\ (2011)]{gary3}
D.~Breton, and et~al.
{\em Nuclear Instruments and Methods}, \textbf{A629}, 123-132, (2011)
\bibitem[Ritt (2008)]{stefan1}
S.~Ritt
{\em IEEE NSS/MIC 2008}, 1512-1515, (2008)
\bibitem[Tescaro (2012)]{stefan2}
D.~Tescaro
{\em IEEE NSS/MIC (2012)}, 1901-1904, (2012)
\bibitem[Paoletti (2013)]{stefan3}
R.~Paoletti
{\em IEEE NSS/MIC (2013)}, 1-4, (2013)
\bibitem[Schellenberg (2014)]{stefan4}
G.~Schellenberg
{\em IEEE NSS/MIC (2014)}, 1-3, (2014)
\bibitem[Ahn et al. (2009)]{cream}
H.~S.~Ahn, and et~al.
{\em Nuclear Instruments and Methods A}, \textbf{602}, 525-536 (2009) 
\bibitem[Abdo et al. (2007)]{milagro}
{Milagro Collaboration}, A.~A. Abdo, and et~al.
{\em Astrophysical Journal Letters}, \textbf{664}, L91-L94 (2007)
\bibitem[Porro et al.\ (2008)]{porro}
M.~Porro, and et~al.
{\em IEEE NSS/MIC (2008)}, N14-7
\bibitem[Abeysekara et al.(2013)]{HAWC}
{HAWC Collaboration}, A.~U. {Abeysekara}, and et~al.
{\em Astroparticle Physics}, \textbf{50--52}, 26--32 (2013)
\bibitem[Vassiliev(1999)]{VERITAS}
{VERITAS Collaboration}, V.~V. {Vassiliev} and et~al. 
{\em Astroparticle Physics}, \textbf{11}, 247--249 (1999)
\bibitem[Achterberg et al.(2006)]{IceCube}
{IceCube Collaboration}, A. {Achterberg}, and et~al.
{\em Astroparticle Physics}, \textbf{26}, 155--173 (2006)
\bibitem[Jetter et al.(2012)]{lognorm}
S.~Jetter et~al.
{\em Chinese Physics C}, \textbf{36(8)}, 733 (2012)
\bibitem[Pivarski(2016)]{pyminuit}
J.~Pivarski.
pyminuit. http://github.com/jpivarski/pyminuit.
\bibitem[Varner (2006)]{gary5}
G.~Varner
{\em JINST}, \textbf{1} P07001 (2006)
\end{thebibliography}
\end{document}